\documentclass{elsart3p}

\usepackage{graphicx}

\usepackage{amssymb}

\begin{document}

\newcommand{\pt}{$p_T$ }
\begin{frontmatter}


\title{Studying Jets with Identified Particles at PHENIX}
\author{Anne Sickles}
\ead{anne@bnl.gov}
\address{Brookhaven National Laboratory, Upton NY 11973}
\author{for the PHENIX Collaboration}

\begin{abstract}
A surprising excess of protons at intermediate $p_T$, 2-5GeV/c, has been observed
in Au+Au collisions at RHIC, for which the source is not known.
In p+p collisions, particles at this $p_T$ arise from jet fragmentation,
however the observed baryon yield in central Au+Au collisions are not compatible
 with
the usual jet fragmentation function.  Two particle $\Delta\phi$
 correlations are a
powerful probe for quantitatively understanding the modifications to jet 
fragmentation from interactions with the medium. Earlier studies 
have shown that the excess baryons do have jet-like partners, 
indicating a hard scattering origin. We present new results from a 
systematic study of two particle correlations 
as a function of trigger and partner particle species, 
charge, $p_T$ and centrality from the high statistics Au+Au dataset.
p+p collisions are also analyzed as a reference.

\end{abstract}

\end{frontmatter}

\section{Introduction}
A remarkable feature of relativistic heavy ion collisions
is the greatly enhanced production of baryons and antibaryons
relative to mesons.  This enhancement over $p+p$
collisions occurs at intermediate transverse momenta (\pt), 
2-5GeV/$c$ \cite{ppg15,starlambda}.  In this momentum
range in $p+p$ collisions particle production shifts from soft, low
momentum transfer processes to hard scattering processes
dominated by jet production followed by fragmentation.
Due to the complicated system in central Au+Au collisions,
it is important to determine whether the baryon and antibaryon
excess is caused by hard or soft processes.  
Previous studies 
have shown that $p$ and $\bar{p}$ between 2.5 and 4.0GeV/$c$ 
are associated with jet-like partners in a 
similar manner as mesons at the same $p_T$ \cite{ppg33},
indicating a hard scattering origin for the baryon and
antibaryon excess.

Models based on 
hadronization by valence quark recombination
have been very successful describing the
particle spectra and elliptic flow at intermediate $p_T$
\cite{friesprc,grecoprc,hwa1,hwa2}.  In the recombination
picture quarks close
together in phase space come together to form
final state hadrons.  In this manner hadrons 
originating from the soft region can dominate in
the region $p_T>$2GeV/$c$ which in p+p collisions
is generally understood to be hard physics.
In general, this mechanism 
enhances baryon production more than meson production at a given
$p_T$ because of the extra $p_T$ in the baryon
from the extra valence quark.

Previous measurements \cite{ppg33} are incompatible
with hadron production at intermediate $p_T$ being dominated by purely 
soft, uncorrelated, recombination.  However,
in some models\cite{grecoprc,hwa1,friescs}, a fraction of  
valence quarks are themselves associated with a
hard scattering leading to modified 
jet fragmentation at intermediate $p_T$.  
The model 
of Fries {\it et al.} \cite{friescs} is the only one to provide
calculations in the \pt range of the data in \cite{ppg33}.
The calculations shows agreement with the data illustrating 
the propagation of hard partons through
the produced dense matter followed by recombination is a
``plausible" mechanism for the origin of the observed correlations 
\cite{friescs}.   

Two particle correlations have been widely used to
study jets in heavy ion collisions 
\cite{starb2b,ppg33,starlowpt,ppg32,stardijet} where, due to the high
multiplicity and moderate jet, energy direct reconstruction 
of jets by standard algorithms is not possible.
In this approach particles are divided into two classes,
{\it triggers} and {\it partners}.
In this work triggers and partners are classified
by their $p_T$, particle type and charge.
 A distribution of the azimuthal angular difference 
$\Delta\phi$ between trigger 
partner pairs is constructed.  The centrality and particle
type dependence of the conditional partner yields per trigger
provide information on the the role of jets in hadron
production at intermediate \pt and  modifications to the fragmentation
process
due to the presence of the medium in Au+Au collisions.

In Reference \cite{ppg33} 
correlations of baryons
($p$,$\bar{p}$) and mesons ($\pi$,$K$) with charged particles
were studied.  The
results presented here extend that work by studying correlations
where both particles are identified.  This provides a model 
independent way to study the effects of the medium on
jets and fragmentation at intermediate \pt.  
The \pt range has been chosen so that the trigger particles
come from a region where the $p/\pi$ ratio at its
maximum and constant, 2.5$<p_T<$4.0GeV/$c$.  This is
the most direct way to experimentally probe the observed
baryon excess.

\section{Experimental Setup}
Charged particles are reconstructed in the central arms of PHENIX 
using drift chambers, each with an azimuthal coverage of $\pi/2$
and one layer of multi-wire proportional chamber with pad readout
(PC1) \cite{trackingnim}.
Here triggers have
2.5$<p_T<$4.0GeV/$c$  and partners with 1.7$<p_T<$2.5GeV/$c$.
Particle identification via time of flight is done for both types
of particles.  The PHENIX high resolution time of flight (TOF) gives
$K/p$ separation to $\approx$4.0GeV/$c$ and is used for the trigger
particle identification.  The PHENIX lead-scintillator electromagnetic
calorimeter (EMCal) provides $K/p$ separation out to $\approx$2.5GeV/$c$.
Partners are identified in either the EMCal or the TOF, which
together cover the entire PHENIX azimuthal acceptance.
For both triggers and partners a 2$\sigma$ match is required between
the track projection and the hit in the particle identification
detector.  
The momentum cuts have been chosen to avoid resonance decays
which could mimic the jet signal.  No correction has been made for
protons originating from the feeddown of $\Lambda$ and $\bar{\Lambda}$.

We perform a correction for PHENIX's non-uniform pair acceptance
in $\Delta\phi$.
This correction is constructed by measuring the 
$\Delta\phi$ distribution from trigger-partner pairs where each
particle is from a different event.  Dividing by this correction
removes the effects of the PHENIX acceptance and leaves
only the true correlations.  The combinatoric
background level from the underlying event multiplicity
is determined absolutely by the convolution of
the trigger and partner single particle rates with
an additional correction for centrality 
correlations \cite{ppg33} which raises the combinatoric
background level by $\approx$0.2\% in the most central collisions
and $\approx$25\% in peripheral collisions.
A correction for the partner efficiency is applied by matching
the observed partner rates to those in \cite{ppg26}.
No extrapolation is made for particles beyond the PHENIX
$|\eta|<0.35$ acceptance.  Such a correction would be
dependent upon an assumption of the shape of the jet profile.
The azimuthal correlation
from elliptic flow is removed by modulating the combinatoric background
level by $1+2v_2^{trig}v_2^{part}\cos(\Delta\phi)$ where
$v_2^{trig}$ and $v_2^{part}$ are the $v_2$ values for
the trigger and partner, respectively, from \cite{ppg22}.
Because the centrality binning in this analysis is finer than
in \cite{ppg22} the $p_T$ integrated centrality dependence is
used to interpolate $v_2$ for collisions more central than
20\%.  

\begin{figure}
\centering
\includegraphics[width=0.5\textwidth]{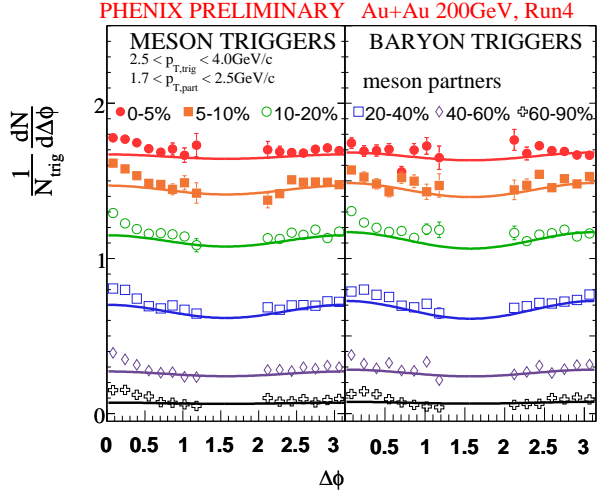}
\caption{$\frac{1}{N_{trig}}\frac{dN}{d\Delta\phi}$ distribution
for meson (left) and baryon (right) triggers, 
2.5$<p_T<$4.0GeV/$c$, and meson partners, 1.7$<p_T<$2.5GeV/$c$,
for six centrality selections in Au+Au.  The solid lines
show the calculated combinatoric background level modulated
by the elliptic flow values.  See text for explanation.}
\label{fig_dphi}
\end{figure}

\section{Results}
Figure \ref{fig_dphi} shows the azimuthal angular difference,
$\Delta\phi$ between trigger mesons (left panel) and baryons (right 
panel) with partner mesons in Au+Au collisions for 
six centralities.  The solid lines show the combinatoric 
background level modulated by the trigger and partner $v_2$
values as described above.  The region around $\Delta\phi=\pi/2$
has very limited pair acceptance due to the requirement that
the trigger particle be identified in the TOF.  For trigger
mesons a near side jet peak is visible.  For trigger baryons
a near side peak is clear for most centralities.

\begin{figure}
\centering
\includegraphics[width=0.5\textwidth]{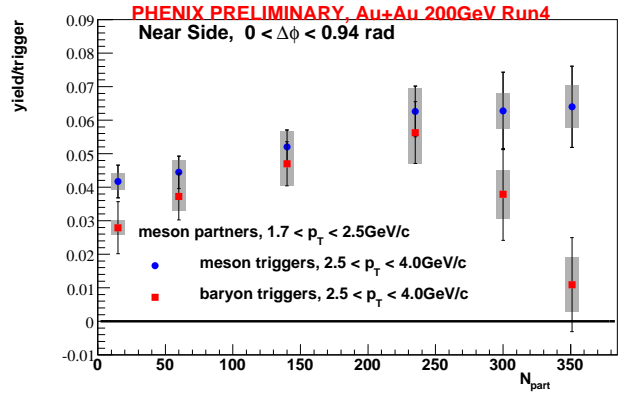}
\caption{Conditional partner yield per trigger for the same
jet, 0$<\Delta\phi<$0.94rad, 
for correlations between baryon,$p$,$\bar{p}$ (squares), and meson,
$\pi$, $K$ (circles), triggers with meson partners. Triggers have
2.5$<p_T<$4.0GeV/$c$ and partners have 1.7$<p_T<$2.5GeV/$c$.}
\label{figmes}
\end{figure}

Figure \ref{figmes} shows that the conditional yield of meson
partners per baryon trigger above the combinatoric background
for $\Delta\phi<$0.94rad.  The systematic
errors, shown as gray boxes, are primarily from
the statistical errors on the $v_2$
values used, the systematic error on the $v_2$
due to the reaction plane resolution
and the correction for centrality correlations.
There is a significant difference in the conditional 
meson yield
between baryon and meson triggers only in the most central 
collisions.  The yield per meson trigger rises slightly
with increasing centrality, while the yield per baryon 
triggers is consistent with the yield per meson trigger
until the most central collisions when it is consistent with
or less than the yield in peripheral collisions.

\begin{figure}
\centering
\includegraphics[width=0.5\textwidth]{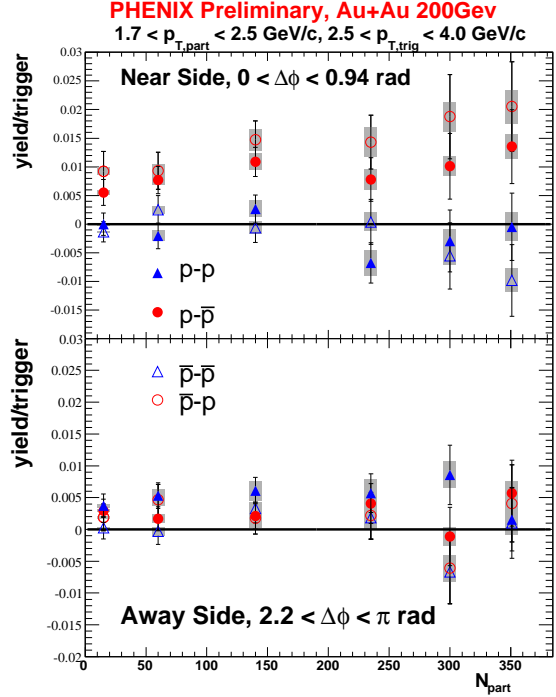}
\caption{Conditional partner yield per trigger for both the 
same side, 0$<\Delta\phi<$0.94rad, (top panel) 
and away side, 2.2$<\Delta\phi<\pi$rad, (bottom panel) for charge
selected correlations between $p$ and $\bar{p}$.  $p$ ($\bar{p}$)
triggered data are filled (hollow) markers.  Circles (triangles)
are for opposite (same) sign pairs. Triggers have
2.5$<p_T<$4.0GeV/$c$ and partners have 1.7$<p_T<$2.5GeV/$c$.}
\label{figppbar}
\end{figure}

Figure \ref{figppbar} shows the number of associated protons or
anti-protons per proton or anti-proton trigger.
on both the near (top panel) and away
side, the regions of near side jet and away side jet fragmentation, respectively
as a function of the number of collision participants.  The yields
are integrated over $\Delta\phi<$0.94 ($\pi-\Delta\phi<$0.94) for
the near (away) side yields.  
The main systematic errors are the same as in Figure \ref{figmes}
The opposite sign pairs have a non-zero
conditional yield and the same sign pairs have a conditional
yield consistent with zero.

\section{Discussion}
There is no significant difference between the baryon
and meson triggered conditional yield of meson 
partners except in the most central collisions in
contrast with the excess 
of protons and antiprotons in the single particles
where the yield, in this same \pt range, scales with
the number of binary nucleon-nucleon collisions
at all centralities \cite{ppg15} while the mesons are
suppressed with increasing centrality.
Therefore, the decrease in the conditional yield
does not lead to the conclusion that 
the excess baryon production at intermediate \pt is due
to purely soft sources.  
A comparison of Figures \ref{figmes} and \ref{figppbar}
shows that in the most central collisions such the probability for a 
baryon triggered jet to contain a meson
in the partner $p_T$ range
is consistent with the probability it contains an oppositely 
charged baryon.
The calculation in \cite{friesmit}, which incorporates
soft parton correlations and jet fragmentation,
does not show this strong centrality dependence for
these correlations.

The lack of centrality dependence in
Figure \ref{figppbar} is interesting in light
of the fact that the $p/\pi^+$ and $\bar{p}/\pi^-$ ratios in the trigger 
and partner $p_T$
ranges increase by about a factor of four in
central Au+Au collisions relative to $p+p$
collisions.  
We do not have p+p data
for comparison at present, but the conditional yield
 results are qualitatively consistent
with PYTHIA \cite{pythia} simulations. 
The lack of centrality dependence to the conditional yield for
both the opposite and same sign pairs is consistent with
$p$ and $\bar{p}$ production in close angle pairs.
While baryon number must be globally conserved,
the novel baryon and antibaryon 
production mechanisms that have been proposed do
not necessarily require the ``extra" $p$ and $\bar{p}$
appear to conserve baryon number locally in a manner
consistent with jet fragmentation, although no theoretical
calculations have been done for this observable.
It should be noted that
the small value of these conditional yield (0.01) does not imply
any ``missing" baryons.  In these conditional yields
the partners are measured in a small $p_T$ window and, 
as noted above, are not corrected
for the partner falling outside the PHENIX psuedorapidity acceptance,
$|\eta|<$0.35. Such a correction would be model dependent and
would be especially unwise given the interesting STAR
data on jet elongation in $\Delta\eta$ \cite{Putschke}.

Jet correlations at intermediate \pt are a powerful probe
of the medium created in heavy ion collisions at RHIC.  
In addition to the particle type dependences shown here a
strong modification to the away side jet shape has been discovered
\cite{ppg32} and elongated correlations in $\Delta\eta$ 
\cite{Putschke}.
Current models are unable
to explain all of these interesting phenomena, but, from the 
data presented here, we know that jet fragmentation on the
near side is modified by the medium produced in central 
Au+Au collisions.  The jets are richer in baryon and 
antibaryon pairs.  In 
the most central collisions baryon triggered jets have significantly 
fewer meson partners than baryon triggered jets in less central collisions
and than meson triggered jets at similar centralities.
Clearly, the picture of jets at intermediate \pt
originating purely from the  surface  of 
the collision region
is inconsistent with these data.  The same side
correlations at intermediate \pt are probing the produced medium 
in Au+Au collisions and a
systematic description of the particle type, centrality and
charge dependence of these correlations are crucial to understanding the 
interactions between hard scattered partons and the medium.

\label{}



\bibliographystyle{elsart-num}
\bibliography{sickles_hp}





\end{document}